\documentclass[aps,twocolumn,superscriptaddress,prd,nofootinbib,nolongbibliography]{revtex4-2}

\usepackage{graphicx}
\usepackage{tabularx}
\usepackage{enumerate}
\usepackage{url}
\usepackage{multirow}
\usepackage{amsmath}
\usepackage{amssymb}
\usepackage{amsfonts}
\usepackage[colorlinks,linkcolor=blue,anchorcolor=blue,citecolor=red]{hyperref}
\usepackage{xcolor}
\usepackage{acronym}
\usepackage{CJK}
\usepackage{orcidlink}

\newcommand{\kiaa}{Kavli Institute for Astronomy and Astrophysics, Peking University, Beijing 100871, China}

\begin{document}
\begin{CJK*}{UTF8}{gbsn}

\title{Constraining Dipole Radiation with Multiband Gravitational Waves from Eccentric Binary Black Holes}

\author{Han Wang (王晗)\orcidlink{0009-0007-5095-9227}}
\email{han.wang.kiaa@pku.edu.cn}
\affiliation{\kiaa}

\author{Lijing Shao (邵立晶)\orcidlink{0000-0002-1334-8853}}
\email{lshao@pku.edu.cn}
\affiliation{\kiaa}
\affiliation{National Astronomical Observatories, Chinese Academy of Sciences, Beijing 100101, China}

\date{\today}

\begin{abstract}
	Dipole-radiation-like deviations from general relativity are most prominent during the early inspiral of compact binaries, making space-ground multiband observations a potential probe of such effects.
	In the same regime, orbital eccentricity can leave a significant imprint on the waveform and is therefore essential for robust dipole-radiation constraints.
	For the first time we present a multiband Bayesian inference pipeline for stellar-mass binary black holes that simultaneously incorporates eccentricity and a theory-agnostic dipole-radiation correction.
	We find strong degeneracies among the dipole parameter, chirp mass, and eccentricity, showing that eccentricity can broaden the inferred dipole posterior by opening an additional degeneracy direction.
	Even so, for a GW231123-like source, one year of TianQin or LISA observation with ground-informed priors from a next-generation detector network can still constrain the dipole parameter to $|b|\lesssim\mathcal{O}(10^{-7})$ under inference with noisy data.
	Our results show that multiband binary black hole observations provide a promising and distinct channel for testing theory-agnostic dipole radiation, while also highlighting the need for more complete waveform modeling in future precision tests of gravity.
\end{abstract}

\maketitle
\end{CJK*}

\acrodef{GW}{gravitational wave}
\acrodef{sBBH}{stellar-mass binary black hole}
\acrodef{EM}{electromagnetic}
\acrodef{SNR}{signal-to-noise ratio}
\acrodef{FIM}{Fisher information matrix}
\acrodef{MCMC}{Markov Chain Monte Carlo}
\acrodef{PSD}{power spectral density}
\acrodef{LISA}{Laser Interferometer Space Antenna}
\acrodef{PN}{post Newtonian}
\acrodef{TDI}{Time delay interferometry}
\acrodef{SPA}{stationary phase approximation}
\acrodef{AGN}{active galactic nuclei}
\acrodef{GR}{general relativity}

\section{Introduction}
Over the past decade, \ac{GW} astronomy has rapidly evolved from first detections to population science \cite{Abbott2016}.
The LIGO-Virgo-KAGRA (LVK) catalogs now contain hundreds of confident compact binary coalescences, and the catalog is ever growing \cite{Abac2025a,Abac2025b}.
Most observed systems are consistent with binaries that have largely circularized before entering the sensitive band of current ground-based detectors, enabling precision measurements of source properties and stringent tests of \ac{GR} \cite{LIGOScientific:2026qni}.
As the sample grows, constraints on possible beyond-\ac{GR} signatures have steadily improved through both event-by-event studies \cite{Abac2026,Saenger2026} and population-level consistency tests \cite{LIGOScientific:2026fcf, LIGOScientific:2026wpt}. 
Multimessenger observations further sharpen such tests: GW170817 and its electromagnetic counterpart provided stringent gravity tests, including bounds relevant to graviton mass and additional radiation channels such as the dipole emission \cite{Abbott2019a, Zhao:2019suc}.

Despite the strong observational preference for quasi-circular signals in current LVK detections, residual orbital eccentricity remains astrophysically plausible.
Eccentricity can be retained in formation channels that differ from isolated binary evolution, including dynamical interactions in dense stellar environments, hierarchical triples, and gas-driven scenarios such as \ac{AGN} disks \cite{Belczynski2016, Barack2019, Breivik2016, Rodriguez2016, Zhang:2019puc, Zhang:2021pwe}.
Therefore, eccentricity provides a valuable probe of a binary's dynamical history and can help distinguish formation mechanisms.

Current ground-based detectors are highly sensitive to the late inspiral-merger-ringdown part of coalescence, and can therefore constrain many intrinsic source parameters with high precision.
However, as the binary approaches merger, gravitational radiation efficiently circularizes the orbit \cite{Peters:1963ux}.
In addition, the in-band observation time is short---seconds to minutes for \acp{sBBH}---which limits access to information about the earlier dynamical state.
As a result, while dedicated searches and population studies have investigated eccentricity signatures, there is so far no unambiguous high-significance evidence that a detected LVK \ac{sBBH} merger contains non-negligible eccentricity \cite{Abbott2019, Abac2024, Xu2026}.
Nevertheless, tentative eccentric features have been reported for several individual \ac{sBBH} events \cite{RomeroShaw2021, Gupte2025, Planas2025}, with GW190521 being a notable candidate for a dynamical origin \cite{RomeroShaw2020, Gamba2023}.
Such cases highlight the potential importance of dynamical formation channels and motivate further studies of eccentric \ac{sBBH} mergers.
Looking ahead, third-generation detectors such as the Einstein Telescope (ET) \cite{Punturo2010} and Cosmic Explorer (CE) \cite{Evans2023, Kalogera:2021bya} will significantly improve sensitivity, particularly at lower frequencies.
This will allow sources to be observed over a longer inspiral and with higher \ac{SNR}, lowering the smallest eccentricity distinguishable from quasi-circular evolution.

Space-based observatories provide a new and complementary window for \ac{GW} observations \cite{AmaroSeoane2017,Luo2016}.
For \acp{sBBH}, a space detector can track inspiral of the system years before merger, where formation-memory features such as eccentricity are less suppressed.
The long baseline and detector orbital modulation can substantially improve the measurement of parameters such as the chirp mass and sky localization, while preserving correlations among source properties \cite{Liu2020,Toubiana2020}.
In several astrophysical scenarios, the binary can retain detectable and astrophysically informative eccentricity in the millihertz-to-decihertz band \cite{Nishizawa2016,Chen2017}.

These considerations also imply a practical requirement: the eccentricity should be modeled explicitly in the data analysis rather than treated as a negligible correction.
Otherwise, forcing quasi-circular templates onto truly eccentric signals can induce non-negligible mismatch.
More importantly, the inferred source parameters may suffer from systematic bias, especially when subtle beyond-\ac{GR} phase effects are simultaneously being constrained.
However, eccentric waveform modeling and Bayesian parameter estimation come with a significant computational cost, due to the long signal duration in the space band, the enlarged parameter space, parameter degeneracies and multimodality in realistic noise, and the requirement of waveform models that remain accurate across detector frequency bands.
This motivates strategies that reduce computational cost and effective dimensionality without sacrificing physical fidelity.

A particularly powerful approach is \ac{GW} multiband detection \cite{Sesana2016,Vitale2016,Klein2022,Toubiana2022}.
A concrete realization of this idea is the archival-search framework \cite{Wong2018,Ewing2021,Wang2024,Wang2026}, in which information from a ground-based detection is used to guide the search for a corresponding sub-threshold signal in archival space-based data.
Ground-based detectors, especially third-generation facilities, can tightly constrain many source parameters near merger.
These measurements can then dramatically reduce the parameter space relevant to the space-based GW analysis.
For Bayesian inference, this information can be incorporated through data-driven priors constructed from the ground-based posterior, thereby shrinking the sampling volume and concentrating computational effort on the subset of parameters that space-based observations improve most.

Beyond computational efficiency, multiband observations broaden the scope of modified-gravity tests.
Eccentric binaries carry richer harmonic structure and phase evolution in the \ac{GW} signal, which  significantly affects how non-\ac{GR} propagation or radiation effects are inferred relative to the standard \ac{GR} orbital dynamics \cite{Roy2026}.
This is particularly relevant for dipole radiation, a generic signature in many beyond-\ac{GR} scenarios where additional degrees of freedom introduce extra radiation channels.
While dipole emission is often discussed in specific theories, its phenomenology can also be treated more generally as a parameterized deviation in the radiation reaction.
In specific scalar-tensor theories without topological terms, the leading dipole term is often controlled by the difference between the effective charges (or sensitivities) of the two bodies, and therefore vanishes for binary black hole systems.
However, binary black hole observations can still be used to constrain more general dipole-radiation terms in theory-agnostic parameterizations, without tying the analysis to a specific underlying model \cite{Barausse:2016eii}.
This is eventually relevant to newly studied cases such as the scalar Gauss-Bonnet gravity \cite{Doneva:2017bvd}.
Previous studies have shown that multiband \ac{sBBH} observations can place very stringent bounds on phenomenological dipole-radiation corrections under quasi-circular assumptions.
Early Fisher-matrix estimates for LISA--ground multiband observations found prospective bounds on the dipole-radiation parameter $b$ at the $\mathcal{O}(10^{-9})$ level for five-year observations \cite{Barausse:2016eii}. 
Comparable bounds were later obtained with a Bayesian LISA--ground analysis using ground-based merger-time information \cite{Toubiana2020a}. 
Fisher-matrix studies of third-generation and decihertz configurations further suggested that still stronger bounds could be achieved in optimistic CE--LISA and CE--DECIGO scenarios \cite{Liu2020a}.
Because both eccentricity and dipole radiation primarily affect the early inspiral of \acp{sBBH}, their interplay is especially important for multiband observations.
Existing constraints on dipole-radiation-like deviations are still largely derived under the quasi-circular assumption, and the extent to which eccentricity reshapes those constraints remains insufficiently explored.

In this paper, for the first time we implement an end-to-end multiband Bayesian framework for inferring dipole radiation in eccentric \acp{sBBH}.
Section~\ref{sec2} presents the signal model, including the eccentric inspiral waveform, the parameterized dipole-radiation correction, and the frequency-domain detector response for space-based detectors.
Section~\ref{sec3} describes the Bayesian inference setup and the detector configurations adopted in our analysis.
Section~\ref{sec4} presents the multiband parameter-estimation results.
Section~\ref{sec5} summarizes the main conclusions and discusses their implications.
We use units where $G=c=1$ unless otherwise specified.

\section{Eccentric waveform model with dipole radiation\label{sec2}}

In this section, we present the signal model adopted for the space-based part of our multiband GW analysis. 
We begin with the \texttt{EccentricFD} description of eccentric inspirals in the frequency domain in Sec.~\ref{sub:sec:ecc}, and then introduce the parameterized dipole-radiation correction used in this work in Sec.~\ref{sub:sec:dipole}.
We next briefly introduce in Sec.~\ref{sub:sec:response} the frequency-domain response of space-based detectors, taking into account the time-dependent \ac{TDI} response for eccentric harmonics. 

\subsection{Eccentric harmonics in GW signals}\label{sub:sec:ecc}

We model the \ac{GW} inspiral waveform using the \texttt{EccentricFD} approximation, a frequency-domain waveform for nonspinning compact binaries \cite{Yunes2009,Huerta2014}.
The model is calibrated for moderate eccentricities, typically $e \lesssim 0.4$.
%In this study, we focus on eccentricity in the space-band inspiral.
The circularization driven by gravitational radiation can be described approximately by \cite{Peters1964,Yunes2009}
\begin{equation}
	e(f) \simeq e_0 \left({f}/{f_0}\right)^{-19/18},
\end{equation}
at leading \ac{PN} order, where $e_0$ is defined at a reference frequency $f_0$.
For example, a binary with $e_0=0.1$ at $f_0=0.01\,\mathrm{Hz}$ is expected to enter the ground-based band ($f \gtrsim 1 \,\mathrm{Hz}$) with $e \lesssim 10^{-3}$.
Thus, a system that is mildly eccentric in the millihertz band can become effectively quasi-circular by the time it reaches the ground-based band.
Motivated by this, we model the binary as eccentric only in the space-based analysis and treat the ground-based signal as quasi-circular.
Because \acp{sBBH} typically merge at frequencies above the most sensitive range of space-based detectors, an inspiral-only, moderate-eccentricity waveform such as \texttt{EccentricFD} is adequate for our purposes.

Eccentric inspirals can be decomposed into harmonics of the orbital motion.
In the frequency domain, and within the stationary-phase approximation, each harmonic can be expressed in terms of the mean orbital frequency $F$.
%For the $j$th harmonic, the \ac{GW} frequency satisfies $f\simeq jF$.
In the \texttt{EccentricFD} approximation \cite{Yunes2009}, the two polarization components take the form
\begin{equation}
	\begin{aligned}
		\tilde{h}_{+,\times}(f)
		= - & \left(\frac{5}{384}\right)^{1/2}\pi^{-2/3}\frac{\mathcal{M}^{5/6}}{D_L} f^{-7/6} \\
		&\times \sum_{j=1}^{10}\xi_{j}^{+,\times}\left(\frac{j}{2}\right)^{2/3}e^{{\rm i}\Psi_{j}(f)} ,
	\end{aligned}
\end{equation}
where $\mathcal{M}=(m_1m_2)^{3/5}(m_1+m_2)^{-1/5}$ is the chirp mass of component masses $m_1$ and $m_2$, and $D_L$ is the luminosity distance.
The coefficients $\xi_{j}^{+,\times}$ depend on the source orientation and evolving eccentricity,
\begin{equation}
	\begin{aligned}
		\xi_{j}^{+,\times}&=\xi_{j}^{+,\times} \left(\iota,\varphi,e(F)\right) \\
		&= \frac{{{{\left( {1 - {e^2}} \right)}^{7/4}}}}{{{{\left( {1 + \frac{{73}}{{24}}{e^2} + \frac{{37}}{{96}}{e^4}} \right)}^{1/2}}}}\left( {C_{ + , \times }^{(j)} + iS_{ + , \times }^{(j)}} \right),
	\end{aligned}
\end{equation}
where $\iota$ is the inclination angle between the orbital angular momentum and the line of sight, and $\varphi$ is the azimuthal phase specifying the direction of \ac{GW} emission in the source frame.
$C_{+,\times}^{(j)}$ and $S_{+,\times}^{(j)}$ are polynomials of $e$, whose coefficients depend on the source orientation through trigonometric functions of $(\iota,\varphi)$ (see Appendix~B in Ref.~\cite{Yunes2009}).
In the quasi-circular limit $e\to 0$, only the $j=2$ harmonic survives, and the coefficients reduce to angular factors of the standard quadrupole (dominant mode) inspiral waveform with
\begin{equation}
	\begin{aligned}
		\xi_{j=2}^+ &= 4 \left( \frac{1+\cos^2\iota}{2}  \right) e^{{\rm i}2\varphi},\\
		\xi_{j=2}^\times &= -4\cos\iota\,e^{{\rm i}2\varphi},\\
		\xi_{j\neq 2}^{+,\times} &= 0.
	\end{aligned}
\end{equation}
The Fourier-domain phase includes \ac{PN} corrections up to 3.5\,PN order.
It can be written as
\begin{equation}
	{\Psi _j} = \frac{\pi }{4} + j{\phi _c} - 2\pi f{t_c} + \frac{3}{{128 x}}{\left( {\frac{j}{2}} \right)^{8/3}}{\Xi _{j}},
\end{equation}
where $x\equiv{\left( {\pi \mathcal{M}f} \right)}^{5/3}$, and $(t_c,\phi_c)$ are the coalescence time and phase.
The explicit expression for $\Xi_j$ is given in Appendix~A of \citet{Huerta2014}.
In \texttt{EccentricFD}, both the amplitude and phase are first expanded in the instantaneous eccentricity $e$ up to $\mathcal{O}(e^8)$, and are then re-expressed in terms of the reference eccentricity $e_0$ and truncated at $\mathcal{O}(e_0^8)$.

To ensure that the \ac{GW} signal is evaluated only within the valid frequency range of each detector, we explicitly bandpass the waveform.
For eccentric inspirals, the mapping between the orbital frequency and the Fourier frequency depends on the harmonic index $j$ (with $f \simeq jF$).
We therefore apply a harmonic-dependent gate to the waveform, i.e. \cite{Arun2007,Yunes2009}
\begin{equation}
	\tilde h_{\rm det}(f)=\tilde h(f) \times \Theta \left(jf_{\rm high}-2f\right)\Theta \left(2f-jf_{\rm low}\right),
\end{equation}
where $\Theta(\cdot)$ denotes the Heaviside step function, and $[f_{\rm low},f_{\rm high}]$ is the detector band in terms of the dominant ($j=2$) GW frequency.
The interval $[f_{\rm low},f_{\rm high}]$ is chosen consistently with the frequency band used in the likelihood evaluation described below.
We also adopt a standardized eccentricity convention by defining the eccentricity at a fixed reference \ac{GW} frequency $f_{\rm ref}=0.01 \, {\rm Hz}$, i.e.\ $e_{\rm ref}\equiv e(f_{\rm ref})$.
The \texttt{EccentricFD} evolution is initialized at $f_{\rm ref}$ and subsequently band-limited using the gate above.
For the lower cutoff we take
\begin{equation}
	f_{\rm low}=\max \left(10^{-4}{\rm Hz},f_{\rm start}\right),
\end{equation}
where we adopt the leading-PN expression for the starting frequency $f_{\rm start}$ over an observation time $T$, which is adequate for \acp{sBBH} in the space band (see Appendix~E in Ref.~\cite{Yunes2009}),
\begin{equation}
	f_{\rm start}=(5/256)^{3/8}\pi^{-1}\mathcal{M}^{-5/8}T^{-3/8}.
\end{equation}
In this study, we set $T=1 \, {\rm yr}$.
The upper cutoff is chosen to remain within the reliable frequency range of each mission,
\begin{equation}
	f_{\rm high}=\min \left(f_{\rm ISCO}, f_{\rm end}\right),
\end{equation}
where $f_{\rm ISCO}={1}/{6^{3/2}\pi M}$, and $M=m_1+m_2$.
We take the mission-dependent cutoff frequency to be
\begin{equation}
	f_{\rm end}^{\rm TianQin}=1\,{\rm Hz}, \quad
	f_{\rm end}^{\rm LISA}=0.1\,{\rm Hz}.
\end{equation}

\subsection{Dipole-radiation correction}\label{sub:sec:dipole}

Beyond the quadrupolar radiation predicted by GR, a broad class of modified gravity theories may admit an additional dipole radiation channel, which is often associated with violations of the strong equivalence principle \cite{Will:2018bme, Barausse:2016eii, Shao:2016ezh}.
In such cases, the total \ac{GW} energy flux of a compact binary can be parametrized in a theory-agnostic way as
\begin{equation}
	\dot{E}=\dot{E}_{\rm GR} \left[1+b \left(\frac{M}{a}\right)^{-1}\right],
\end{equation}
where $\dot{E}_{\rm GR}$ is the leading quadrupole flux in GR, $a$ is the semi-major axis of orbit, and $b$ is a dimensionless coefficient characterizing the strength of the dipole correction. 
For a given underlying theory, $b$ can be expressed in terms of the corresponding coupling constants or effective charges of the binary. 
For instance, in Jordan-Fierz-Brans-Dicke-like scalar-tensor theories one has $b=5(\Delta\alpha)^2/96$, where $\Delta\alpha$ denotes the difference between the scalar charges of the two bodies. 
Using $M/a \sim v^2$, with $v$ the characteristic orbital velocity, one finds that the additional term scales as $v^{-2}$ relative to the GR quadrupole flux. 
It therefore corresponds to a relative $-1$\,PN correction, and is expected to be important for widely separated binaries, i.e. in the low-frequency regime.

For eccentric binaries, the orbit-averaged energy flux acquires additional eccentricity-dependent enhancement factors. 
At leading order, the quadrupolar contribution in GR and the dipole correction take different forms in the eccentricity, reflecting the distinct multipole structure of the radiation. 
Adopting the leading-order orbit-averaged results, the total flux can be written as \cite{Will1989,Ma2019}
\begin{equation}
	\begin{aligned}
		\dot{E}^{\rm ecc}=&-\frac{32}{5}\frac{M^5\eta^2}{a^5}\frac{1+\frac{73}{24}e^2+\frac{37}{96}e^4}{(1-e^2)^{7/2}} \\ &-\frac{32}{5}\frac{M^5\eta^2}{a^5} b\left(\frac{M}{a}\right)^{-1}\frac{1+\frac{1}{2}e^2}{(1-e^2)^{5/2}},
	\end{aligned}
\end{equation}
where $ \eta  \equiv \left( {{m_1}{m_2}} \right){\left( {{m_1} + {m_2}} \right)^{ - 2}} $ is the symmetric mass ratio.
Combining this expression with the balance equations and Keplerian orbital dynamics then yields the modified evolution equations for the orbital frequency $F$ and eccentricity $e$, which subsequently enter the waveform through the corrected relation $e(F)$, and hence affect both the harmonic phase and amplitude in the frequency domain.

In the frequency domain, the dipole contribution enters the waveform through the modified inspiral dynamics. 
Because the dipole flux changes the secular evolution of the binary, it affects both the phase accumulation of each harmonic and the normalization of the corresponding amplitude. 
Accordingly, these two effects are incorporated in the model as a correction to $\Xi_j$ in the harmonic phase and an additional dipole-dependent prefactor in the amplitude \cite{Ma2019}.

For the phase sector, we write
\begin{equation}
	\Xi_j = \Xi_j^{\rm GR} + \Xi_j^{\rm dip},
\end{equation}
where the dipole contribution takes the form
\begin{equation}
	\Xi_j^{\rm dip}	= \frac{1}{2}\bar b\left(\frac{2\pi f_0 \mathcal M}{j}\right)^{-2/3}\mathcal F(\chi,e_0),
\end{equation}
with
\begin{equation}
	\chi \equiv \frac{F}{F_0}=\frac{f}{f_0},	\qquad	\bar b \equiv b\eta^{2/5},
\end{equation}
with $\mathcal F(\chi,e_0)$ denoting the corresponding eccentric expansion in powers of $e_0$.
The explicit coefficients are given in Eq.~(37) of Ref.~\cite{Ma2019} and are not repeated here for brevity.

For the amplitude sector, we include the corresponding dipole correction through an overall multiplicative factor,
\begin{equation}
	\xi_{j}^{+,\times} \rightarrow \xi_{j}^{+,\times}\left[1+\bar b \left(\frac{2\pi f \mathcal M}{j}\right)^{-\frac{2}{3}}\frac{(1+\frac{1}{2}e^2)(1-e^2)}{1+\frac{73}{24}e^2+\frac{37}{96}e^4}\right]^{-\frac{1}{2}} .
\end{equation}
In practice, we retain only the leading dipole contribution, namely expand consistently to the linear order in $b$, and truncate the eccentric expansion consistently at the same order as in the baseline eccentric waveform model.

\subsection{Frequency-domain detector response}\label{sub:sec:response}

For space-based detectors, the response is intrinsically time dependent because the constellation moves during the observation.
To suppress laser phase noise, we adopt \ac{TDI} combinations \cite{Tinto:2004wu}.
Starting from the Michelson channels $(X,Y,Z)$, which are constructed from delayed single-link responses, one has,
for example,
\begin{equation}
	\tilde{X}=(1-z^2)\left(\tilde{y}_{31}+z\tilde{y}_{13}-\tilde{y}_{21}-z\tilde{y}_{12}\right),
\end{equation}
where $\tilde{y}_{sr}(f)$ denotes the frequency-domain single-link response for the link from spacecraft $s$ to spacecraft $r$, $z\equiv \exp(i2\pi fL)$ for armlength $L$, and $\tilde{Y}$ and $\tilde{Z}$ follow by cyclic permutation.
Quantities with a tilde denote their frequency-domain representations.
The orthogonal \ac{TDI} channels $(A,E,T)$ are then defined as
\begin{equation}
	\begin{aligned}
		A &= \frac{1}{\sqrt{2}}(Z-X),\\
		E &= \frac{1}{\sqrt{6}}(X-2Y+Z),\\
		T &= \frac{1}{\sqrt{3}}(X+Y+Z).
	\end{aligned}
\end{equation}
We adopt the corresponding channel \acp{PSD} from \citet{Babak2021}.

To evaluate the response in the frequency domain, we follow the prescription of Refs.~\cite{Marsat2018,Marsat2021}.
In this approach, the detector response is evaluated at an effective time associated with each Fourier frequency.
In that framework, the time-frequency correspondence is applied mode by mode to the spin-weighted spherical-harmonic waveform components.
For \texttt{EccentricFD}, however, the natural decomposition is in terms of eccentric harmonics labeled by $j$.
We therefore apply the same construction harmonic by harmonic, with \cite{Wang2026}
\begin{equation}
	t_f^j = \frac{1}{{2\pi }}\frac{{d{\Psi _j}}}{{df}}.
\end{equation}
The \ac{GW} signal can be described as
\begin{equation}
	\begin{aligned}
		H &=\sum\limits_{j} h_j,\\
		{\tilde h}_j &={{P_ + }\tilde h_j^ +  + {P_ \times }\tilde h_j^ \times },
	\end{aligned}
\end{equation}
where $P_+$ and $P_\times$ are the polarization tensors.
The corresponding single-link response is
\begin{equation}
	{{\tilde y}_{\rm sr}} = \sum\limits_j {\mathcal{T}_{\rm sr}^j(f):{\tilde h}_j},
\end{equation}
with the transfer function
\begin{equation}
	\mathcal{T}_{\rm sr}^j(f) = {G_{\rm sr}}\left( {f,t_f^j} \right),
\end{equation}
where
\begin{equation}\label{gsr_e}
	\begin{aligned}
		{G_{\rm sr}}(f,t) =& \frac{{{\rm i}\pi fL}}{2}\operatorname{sinc} \big[ {\pi fL\left( {1 - \mathbf{k} \cdot \mathbf{n}_l} \right)} \big] \\
		&\times \exp \big[ {{\rm i}\pi f\left( {L + \mathbf{k} \cdot \left( {\mathbf{p}_r + \mathbf{p}_s} \right)} \right)} \big] \mathbf{n}_l \otimes \mathbf{n}_l .
	\end{aligned}
\end{equation}
Here $\mathbf{k}$ is the wave vector, $\mathbf{n}_l$ is the unit vector along the link from spacecraft $s$ to $r$, and $\mathbf{p}_s$ and $\mathbf{p}_r$ are the spacecraft positions evaluated at the same time $t$.

In this work, we use the space-detector response as implemented in \texttt{GWSpace} \cite{Li2025}.
The corresponding eccentric waveform, including the dipole correction, is interfaced with \texttt{PyCBC} \cite{Biwer2019} through a waveform plugin for parameter inference.

\begin{table*}
	\centering
	\caption{Injected values and priors for the GW231123-like source used in the ground- and space-based analyses. 
		The quoted ground-based posteriors give the median and 90\% credible intervals and are listed for reference only. 
		The space priors for shared parameters are constructed from a kernel density estimate (KDE) of the ground-based posterior samples. 
		Equatorial sky coordinates $(\alpha,\delta)$ are transformed into ecliptic coordinates $(\lambda,\beta)$ for the space-based analysis. 
		The prior range for $b$ is chosen using the GW170817 constraint as a reference scale, and a two-sided prior is adopted (see main text).
		We use $\mathcal{U}(\cdot)$ for uniform distribution, $\mathcal{U}_{\rm vol}(\cdot)$ for a distribution uniform in volume, and $\mathcal{U}_{\rm sky}$ for an isotropic sky distribution.}
	\renewcommand{\arraystretch}{1.66}
	%\begin{tabular}{c|c|c|c|c} 
	\begin{tabularx}{\textwidth}{p{2.8cm}p{2.2cm}p{3.1cm}p{4.8cm}p{5.1cm}}
		\hline\hline
		                                & Injection     & Ground Prior                                & Ground Posterior (90\% CI)       & Space Prior                                              \\
		\hline%\hline
		%$ m_1 $ [$ M_\odot $]           & 190.43        & -                                           & $190.3020$                       & -                                                        \\
		%$ m_2 $ [$ M_\odot $]           & 140.39        & -                                           & $140.4754$                       & -                                                        \\
		$ \mathcal{M} $ [$ M_\odot $]   & 142.0118      & $\mathcal{U}(141, 143)$                     & $142.010^{+0.023}_{-0.024}$      & \multirow{9}[0]{*}{[KDE from ground posterior]}          \\
		$ q $                           & 1.3564        & $\mathcal{U}(1, 2)$                         & $1.3547^{+0.0068}_{-0.0069}$     &                                                          \\
		$ D_L $ [Gpc]                   & 2.2           & $\mathcal{U}_{\rm vol}(1.5, 3.0)$           & $2.192^{+0.014}_{-0.014}$        &                                                          \\
		$ \alpha $ (RA) [rad]           & 2.9427        & \multirow{2}[0]{*}{$\mathcal{U}_{\rm sky}$} & $2.94274^{+0.00064}_{-0.00065}$  &                                                          \\
		$ \delta $ (DEC) [rad]          & $-0.3431$     &                                             & $-0.34361^{+0.00087}_{-0.00088}$ &                                                          \\
		%$ \lambda $ [rad]               & 3.1016        & -                                           & $3.1018$                         &                                                          \\
		%$ \beta $ [rad]                 & $-0.3927$     & -                                           & $-0.3931$                        &                                                          \\
		$ \Delta t_c $ [ms]             & 0             & $\mathcal{U}(-1, 1)$                        & $3.6^{+4.3}_{-4.5}\times10^{-2}$ &                                                          \\
		$ \phi_c $ [rad]                & 4.7124        & $\mathcal{U}(0, 2\pi)$                      & $4.721^{+0.021}_{-0.022}$        &                                                          \\
		$ \iota $ [rad]                 & 0.5236        & ${\rm Sine}(0, \pi)$                        & $0.529^{+0.011}_{-0.011}$        &                                                          \\
		$ \psi $ [rad]                  & 1.5708        & $\mathcal{U}(0, \pi)$                       & $1.557^{+0.017}_{-0.017}$        &                                                          \\
		$ e_{0.01 \, \mathrm{Hz}} $     & 0 $\big|$ 0.1 & N/A                                         & N/A                              & $\mathcal{U}(0, 0.02)$ $\big|$ $\mathcal{U}(0.09, 0.11)$ \\
		\hline
		$ b $                           & 0             & N/A                                         & N/A                              & $\mathcal{U}(-0.000012, 0.000012)$                       \\
		\hline%\hline
	%\end{tabular}
	\end{tabularx}
	\label{prior}%
\end{table*}                    

\section{Bayesian parameter estimation setup\label{sec3}}

To assess the measurability of dipole-radiation effects, we perform Bayesian parameter estimation for both the ground- and space-based observations.
Throughout this work, the detector data are modeled as the sum of a single \ac{GW} signal, $h(t,\lambda^\mu)$, and stationary, Gaussian noise, $n(t)$,
\begin{equation}
	D(t)=h(t,\lambda^\mu)+n(t),
\end{equation}
where $\lambda^\mu$ denotes the source parameters.

Given data $D$, inference is based on the posterior distribution
\begin{equation}
	P(\lambda^\mu\mid D)
	=\frac{P(D\mid\lambda^\mu)P(\lambda^\mu)}{P(D)}
	\propto \mathcal{L}(\lambda^\mu)P(\lambda^\mu),
\end{equation}
where $P(\lambda^\mu)$ is the prior, $\mathcal{L}(\lambda^\mu)\equiv P(D\mid\lambda^\mu)$ is the likelihood, and the evidence
\begin{equation}\label{evidence}
	P(D)=\int P(D\mid\lambda^\mu)P(\lambda^\mu){\rm d}\lambda^\mu
\end{equation}
acts as an overall normalization constant.
Under the Gaussian-noise assumption, the likelihood takes the standard form
\begin{equation}\label{likelihood}
	\begin{aligned}
		\ln \mathcal{L}&\propto -\frac{1}{2} \big(D-h(\lambda^\mu)\mid D-h(\lambda^\mu) \big)\\
		&=-\frac{1}{2}\big[(D\mid D)+(h\mid h)-2(h\mid D)\big],
	\end{aligned}
\end{equation}
where the inner product is defined by \cite{Finn:1992wt}
\begin{equation}
	(h \mid g)\equiv 4 \Re \int_0^{\infty} \frac{\tilde{g}^*(f) \tilde{h}(f)}{S_n(f)} \mathrm{d} f,
\end{equation}
with $\tilde h(f)$ the Fourier-domain waveform and $S_n(f)$ the one-sided noise \ac{PSD}.

The source modeling differs between the ground- and space-based analyses.
For the ground-based analysis, we use the frequency-domain waveform model \texttt{IMRPhenomXPHM} \cite{Pratten2021} for both signal injection and recovery.
This model describes quasi-circular, precessing binary black holes and includes higher-order multipoles.
Since the binaries considered here are expected to be effectively circular by the time they enter the ground-based band, we neglect orbital eccentricity in the ground analysis.
We also omit the dipole-radiation parameter $b$ in the ground-based analysis.
Although leading dipole-like phase corrections can be incorporated in simpler quasi-circular inspiral waveforms, a consistent implementation in the precessing, higher-mode \texttt{IMRPhenomXPHM} model used here is nontrivial and beyond the scope of this work.
The ground-based analysis is used mainly to provide informed priors on the shared source parameters for the space-based inference, where the low-frequency inspiral carries the primary sensitivity to $b$.
In addition, for consistency with the non-spinning \texttt{EccentricFD} waveform adopted in the space-based analysis, we neglect component spins throughout.
The ground-based parameter set is therefore taken to be
$$ \lambda^\mu_{\rm g}=\left(\mathcal{M},q,D_L,\alpha,\delta,t_c,\phi_c,\iota,\psi\right), $$
where $q\equiv m_1/m_2\geq 1$ is the mass ratio, $(\alpha,\delta)$ are the right ascension and declination in equatorial coordinates, $\iota$ is the inclination angle, and $\psi$ is the polarization angle.
For the space-based analysis, we extend the parameter set by including the reference eccentricity and the dipole-radiation parameter, and we adopt ecliptic sky coordinates. Therefore, we have
$$ \lambda^\mu_{\rm s}=\left(\mathcal{M},q,D_L,\lambda,\beta,t_c,\phi_c,\iota,\psi,e_{0.01\,\mathrm{Hz}},b\right), $$
where $(\lambda,\beta)$ denote the ecliptic longitude and latitude in the Solar-system barycenter frame.

We consider a third-generation network consisting of one ET and two CE detectors for the ground-based analysis.
We use the standard ET triangular design with three co-located $10\,\mathrm{km}$ interferometers \cite{Punturo2010}, and two CE sites based on the $40\,\mathrm{km}$ and $20\,\mathrm{km}$ detector concepts \cite{Evans2023}.
These detector configurations are adopted for simulation purposes only and should not be interpreted as the final detector deployment.
The low-frequency cutoffs are set to $f_{\rm low}=2 \, \mathrm{Hz}$ for ET and $f_{\rm low}=5 \, \mathrm{Hz}$ for CE.
In the space-based case, we consider TianQin and LISA, assuming a continuous one-year observation before binary coalescence and neglecting duty-cycle effects.
The detector noise models follow Ref.~\cite{Luo2016} for TianQin and Ref.~\cite{Babak2021} for LISA, while the corresponding orbital configurations are taken from Ref.~\cite{Hu2018} and Ref.~\cite{Cornish2003}, respectively.

The full inference pipeline is implemented within the \texttt{PyCBC Inference} framework \cite{Biwer2019}.
Posterior sampling is performed with \texttt{nessai}, a nested-sampling algorithm enhanced by normalizing flows to improve sampling efficiency \cite{Williams2021,Williams2023}.
For the long-duration space-based signals considered in this work, direct likelihood evaluation with full-resolution waveforms would be computationally prohibitive.
We therefore adopt a heterodyned likelihood scheme, also referred to as the relative-binning approximation.
In this approach, a single high-resolution reference waveform is computed once, and the likelihood for candidate waveforms is subsequently evaluated by interpolating their ratio to the reference over a much coarser frequency grid \cite{Cornish2010,Zackay2018,Leslie2021}.

It should be noted that different waveform families are used in the ground- and space-based bands, and the two descriptions are therefore not fully self-consistent.
Recent work has made important progress toward more complete eccentric-spin waveform models, including precessing-eccentric inspiral models \cite{Klein2021,Arredondo2024,Morras2025} and EOB-based eccentric coalescence models \cite{Liu2024,Gamboa2025}.
However, constructing a unified model that is simultaneously suitable for long-duration space-band evolution, ground-band merger-ringdown analysis, and the phenomenological dipole-radiation correction considered here remains technically challenging \cite{Thomas2026} and is not currently available for the parameter-estimation setup used in this work.
The use of different waveform families is therefore motivated by waveform suitability and computational cost in the present multiband analysis, while constructing such a unified model is left to future study.

\section{Multiband constraints on dipole radiation\label{sec4}}

In this section, we present the multiband parameter-estimation results for a GW231123-like \ac{sBBH} source.
We first analyze the recovery with a next-generation ground-based network, whose posterior provides the prior information for the subsequent space-based archival inference.
We then study the constraints from inspiral in the space band, comparing different detector combinations as well as circular and eccentric injections.

The injected values and prior ranges of various parameters are listed in Table~\ref{prior}.
The component masses and luminosity distance are taken from Ref.~\cite{Abac2025a}, while the remaining parameters are chosen randomly.
For reference, Table~\ref{prior} also reports the 90\% credible intervals (CIs) from the ground-based posterior.
The source is extremely loud in the next-generation network, with a \ac{SNR} of about 1200, so its parameters are tightly constrained.
As found in the previous multiband studies \cite{Wang2026,Klein2022}, most shared parameters are therefore already well measured on the ground, and the space-based analysis mainly contributes information about the mass parameters \cite{Lyu2023,Sesana2016}.
In the subsequent space-based analysis, the priors on the shared parameters are constructed directly from the ground-based posterior samples using a Gaussian-kernel density estimate, which is adequate for the smooth and unimodal posteriors encountered here.
The 90\% CIs quoted in Table~\ref{prior} are included only as a compact summary and are not used directly as priors.
In a realistic archival search for future events, the simulated ground posterior used here should be replaced by the published posterior samples of the corresponding event.
Using such ground-based posterior information also gives a realistic estimate of the parameter-space volume that must be explored in the space-based inference.

We then perform the space-based Bayesian analysis for two representative injections with $e_{0.01 \, \mathrm{Hz}}=0$ and $e_{0.01 \, \mathrm{Hz}}=0.1$, in order to illustrate the impact of eccentricity.
For the heterodyned likelihood, the reference waveform is generated at the median values of the ground-informed parameters (in Table~\ref{prior}), while the parameters not constrained by the ground analysis ($e_{0.01 \, \mathrm{Hz}}$ and $b$) are set to their injected values.
For the dipole-radiation parameter $b$, we adopt a prior range motivated by existing \ac{GW} constraints, using the GW170817 bound only as a numerical reference scale \cite{Abbott2019a}.
This helps reduce the parameter space to be explored in the Bayesian analysis, while remaining broader than the posterior range obtained below.
We emphasize that this is used only to set a reasonable phenomenological prior range, rather than to transfer the GW170817 constraint directly to \ac{sBBH} systems, since a different source class (here neutron stars versus black holes) and frequency regime may realize modified-gravity effects differently \cite{Barausse:2016eii}.
Because of the constraining ability from space-based data, a broader range of prior on $b$, say, from sBBH constraints, will not likely  change the results in the study.
We further adopt a two-sided prior on $b$, treating it as a phenomenological parameter that describes an effective dipole-radiation-like correction to the inspiral, rather than as a strictly positive amplitude of an additional outgoing flux in a specific gravity theory.

\begin{figure}
	\centering
	\includegraphics[width=\linewidth]{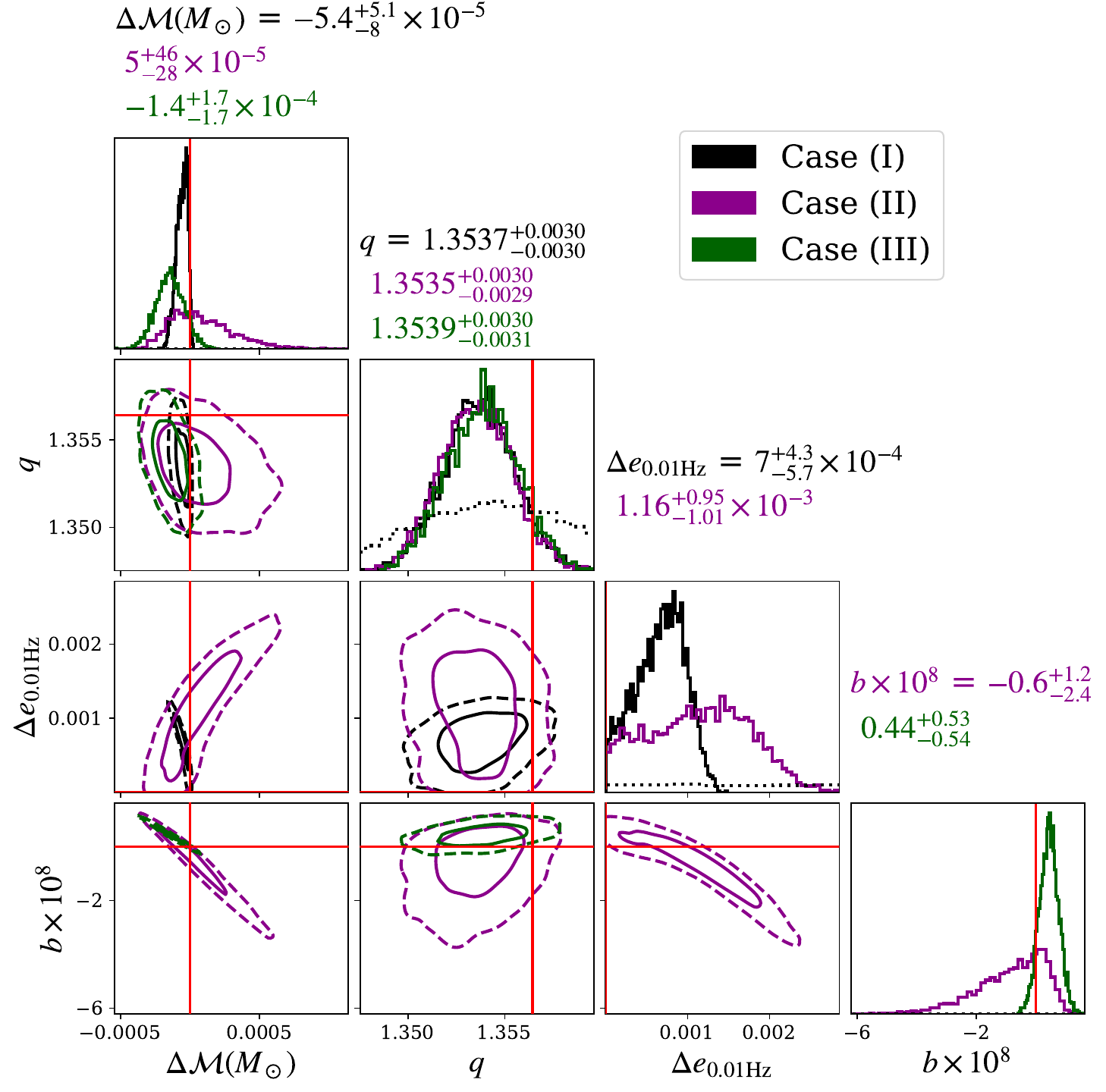}
	\caption{Posterior distributions for four selected parameters $(\mathcal{M}, q, e_{0.01 \,\mathrm{Hz}}, b)$ from the TianQin+LISA Bayesian analysis for the $e_{0.01 \, \mathrm{Hz}}=0$ injection, plotted using $(\Delta\mathcal{M}, q, \Delta e_{0.01,\mathrm{Hz}}, b\times10^8)$, where $\Delta\mathcal{M}=\mathcal{M}-\mathcal{M}^{\rm inj}$ and $\Delta e_{0.01,\mathrm{Hz}}=e_{0.01,\mathrm{Hz}}-e_{0.01,\mathrm{Hz}}^{\rm inj}$.
	Three sampling setups are compared: Case (I) with $b$ excluded (black), Case (II) where both $b$ and $e_{0.01\,\mathrm{Hz}}$ are sampled (purple), and Case (III) with a fixed $e_{0.01 \, \mathrm{Hz}}=0$ (green).
	The red lines indicate the injected values in the plotted coordinates.
	The two-dimensional contours enclose 50\% and 90\% of the posterior probability.
	The dashed curves in the one-dimensional panels show the corresponding prior distributions for reference.}	
	\label{pe_0}
\end{figure}

\begin{figure}
	\centering
	\includegraphics[width=\linewidth]{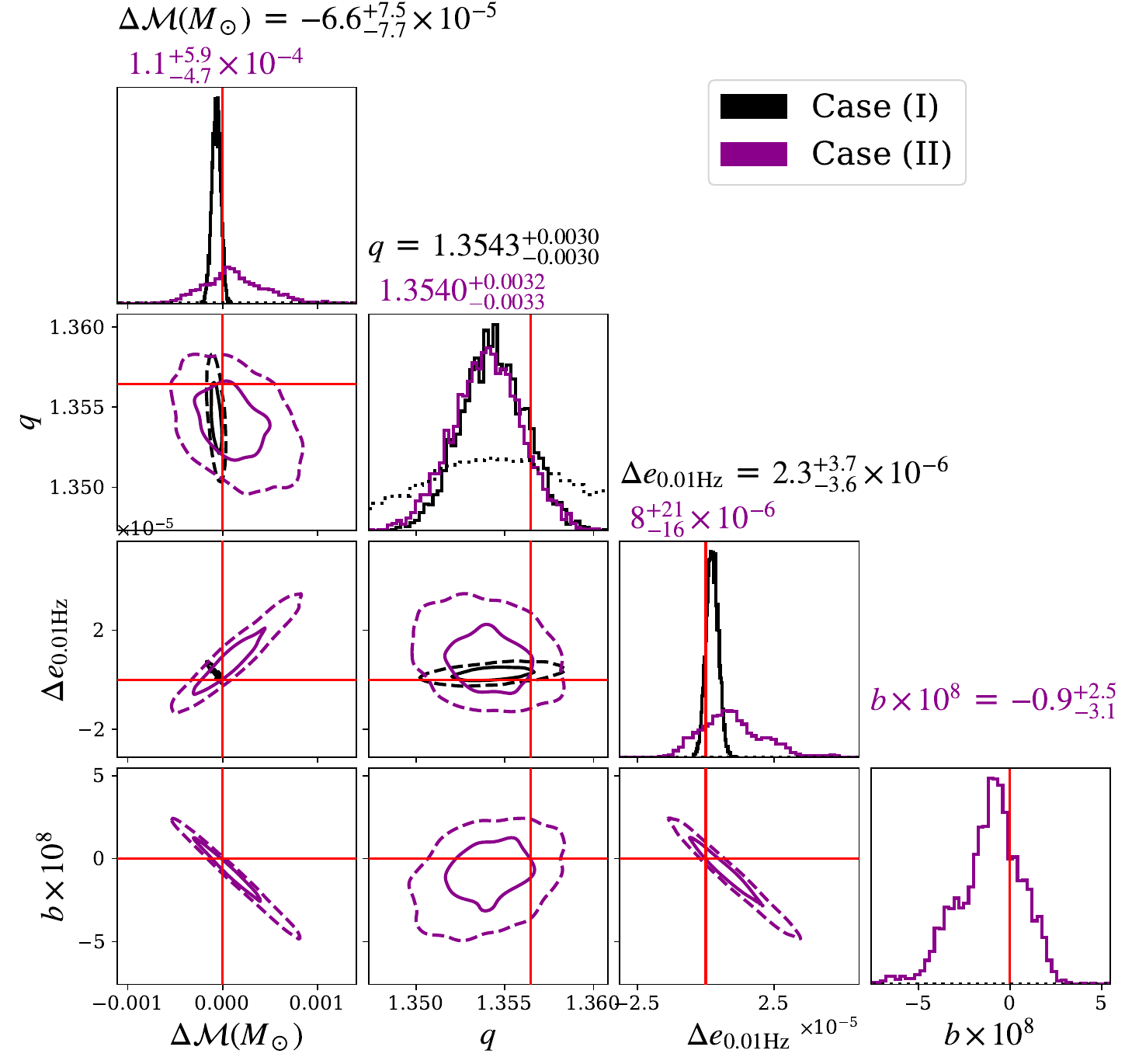}
	\caption{Same as Fig.~\ref{pe_0}, but for the $e_{0.01 \, \mathrm{Hz}}=0.1$ injection, where only Cases (I) and (II) are shown.}
	\label{pe_01}
\end{figure}

Figure~\ref{pe_0} and Figure~\ref{pe_01} show posteriors from selected parameters in the Bayesian analysis for the joint TianQin+LISA observation.
In the posterior figures below, we use recentered or rescaled plotting coordinates, such as $\Delta \mathcal{M}$, $\Delta e_{0.01{\rm Hz}}$, and $b\times10^8$, only for visual clarity.
These transformations do not change the physical parameter correlations discussed in terms of $\mathcal{M}$, $e_{0.01 \,\mathrm{Hz}}$, and $b$.
Although the inference is carried out in the archival search parameter space with all parameters, here we focus on the four parameters most relevant to the dipole-radiation test, i.e. $\mathcal{M}$, $q$, $e_{0.01 \,\mathrm{Hz}}$, and $b$.
The remaining parameters are already tightly constrained by the ground-based prior and show only limited further improvement in the space-based analysis, and exhibit little correlation with $b$. 
We therefore highlight these four dimensions here.
For completeness, the full posterior distributions for the $e_{0.01 \, \mathrm{Hz}}=0.1$ case with TianQin, LISA, and TianQin+LISA are provided in Appendix~\ref{app}.

For the circular injection, we compare three cases: 
\begin{enumerate}[(I)]
	\item a reference analysis without dipole radiation,
	\item  a full dipole analysis in which both $e_{0.01 \, \mathrm{Hz}}$  and $b$ are sampled, 
	\item a restricted dipole analysis in which $b$ is sampled while $e_{0.01 \, \mathrm{Hz}}$ is fixed to zero.
\end{enumerate}
For the eccentric injection, only the first two cases are considered, since an injected eccentricity of $e_{0.01 \, \mathrm{Hz}}=0.1$ cannot be consistently treated as circular.

Case (I) provides the baseline constraint from standard parameter estimation.
In this case, both injections yield tight posteriors:
the chirp mass is measured at $\sim \mathcal{O}(10^{-5}) \, M_\odot$, while the eccentricity is constrained to the order of $\mathcal{O}(10^{-4})$ for the $e_{0.01 \, \mathrm{Hz}}=0$ injection and $\mathcal{O}(10^{-6})$ for the $e_{0.01 \, \mathrm{Hz}}=0.1$ injection.
Once the dipole parameter $b$ is introduced, the posteriors of both $\mathcal{M}$ and $e_{0.01\mathrm{Hz}}$ broaden by roughly an order of magnitude, whereas the mass-ratio posterior remains almost unchanged.
This difference reflects the structure of the parameter degeneracies rather than a trivial increase in the dimensionality of parameter space.
In Case (I), $\mathcal{M}$ and $e_{0.01 \, \mathrm{Hz}}$ already show a clear anticorrelation, consistent with previous studies of eccentric \acp{sBBH}.
Allowing $b$ to vary introduces additional correlations with both $\mathcal{M}$ and $e_{0.01 \, \mathrm{Hz}}$, indicating that all three parameters affect the accumulated early-inspiral phasing in a degenerate way.
The $\mathcal{M}$--$b$ correlation is particularly strong and reshapes the projected $\mathcal{M}$--$e_{0.01 \, \mathrm{Hz}}$ posterior so that its apparent correlation direction can even reverse.
By contrast, $q$ remains largely decoupled from $b$ across all three cases.

The restricted Case (III) analysis with $e_{0.01 \, \mathrm{Hz}}$ fixed to zero helps isolate the specific role of eccentricity in this parameter degeneracy structure.
Even in this case, allowing $b$ to vary still broadens the chirp-mass posterior, showing that $b$ alone already produces a substantial degeneracy with $\mathcal{M}$.
The broadening is nevertheless weaker than in the full analysis in Case (II):
for the circular injection, allowing eccentricity to vary increases the width of the credible interval of $b$ by a factor of approximately three relative to Case (III).
This comparison quantitatively illustrates the additional, non-negligible degeneracy direction introduced by eccentricity.
In other words, a signal with a relatively large $b$, a small eccentricity, and  a small chirp mass can mimic a \ac{GW} with a smaller $b$ but larger eccentricity and chirp mass.
Eccentricity is therefore an essential parameter in carrying out precision tests of dipole-radiation-like deviations from \ac{GR}.

The posterior structure seen in Figs.~\ref{pe_0} and \ref{pe_01} can be understood directly from the phase dependence of the waveform.
The chirp mass $\mathcal{M}$ controls the overall inspiral phasing, i.e.\ it sets the main phase scale of the signal.
The leading \ac{GR} term can be written as
\begin{equation}
	\Psi_{\rm GR}(f) \propto (\pi \mathcal{M} f)^{-5/3}.
\end{equation}
On top of this, both dipole radiation and eccentricity introduce additional corrections that are most important in the early, low-frequency inspiral.
At leading order, the dipole contribution enters as a $-1$\,PN term,
\begin{equation}
	\Delta \Psi_b(f) \propto b\eta^{2/5}(\pi \mathcal{M} f)^{-7/3},
\end{equation}
while the leading eccentric correction is also enhanced at low frequencies via,
\begin{equation}
	\Delta \Psi_e(f) \propto e_0^2 f^{-19/9}\Psi_{\rm GR}(f).
\end{equation}
As a result, changes in $b$ and $e_{0.01 \, \mathrm{Hz}}$ can both be partially absorbed by shifts in the overall phase scale set by $\mathcal{M}$ over a finite observation band, so that producing the correlated posteriors seen among $b$, $\mathcal{M}$, and $e_{0.01 \, \mathrm{Hz}}$.
By contrast, the dependence on the mass ratio is distributed across multiple \ac{PN} contributions and does not introduce a comparably distinct low-frequency phase structure, so its correlation with $b$ remains much weaker.

\begin{figure}
	\centering
	\includegraphics[width=\linewidth]{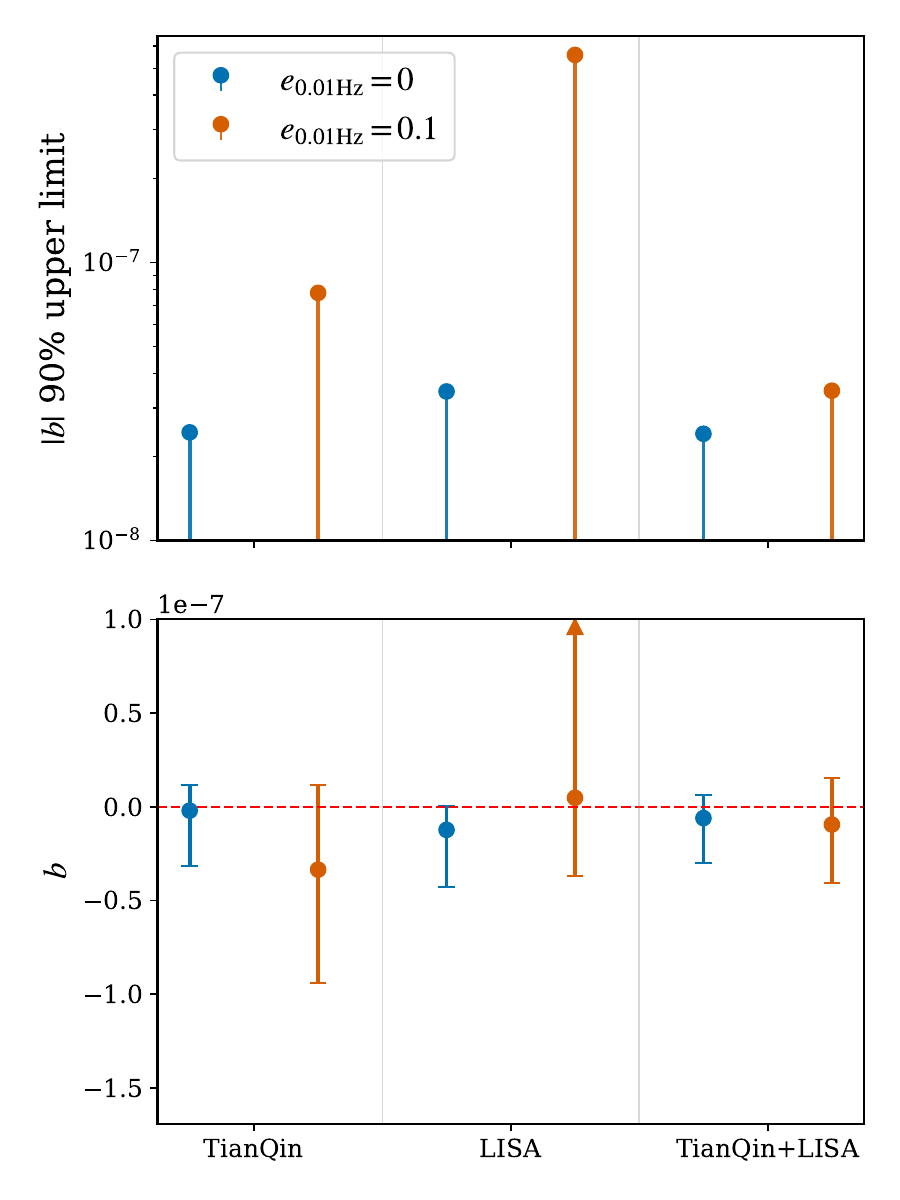}
	\caption{Comparison of dipole-radiation constraints for different space-based detector configurations and injected eccentricities in the multiband analysis of a GW231123-like source with informed priors from ground-based detectors.
	The bottom panel shows the posterior median and 90\% CIs for the dipole parameter $b$.
	The top panel shows the corresponding 90\% upper bound on $|b|$.
	Blue and orange symbols correspond to injections with $e_{0.01 \, \mathrm{Hz}}=0$ and $e_{0.01 \, \mathrm{Hz}}=0.1$, respectively.
	The upward triangle indicates that the upper edge of the 90\% CI extends beyond the plotted vertical range (see main text).}
	\label{b_summary}
\end{figure}

We summarize the corresponding constraints on $b$ for different space-based configurations in Fig.~\ref{b_summary}.
The lower panel shows the posterior median and 90\% CIs of the signed parameter $b$, while the upper panel gives the 90\% upper bound on $|b|$ derived from the same posterior samples.
The combined TianQin+LISA network yields tighter constraints than either detector alone.
This is consistent with its higher network \ac{SNR}, which is $\sim 11$ for the injection considered here, compared with $\sim 7.5$ for TianQin or LISA individually.
In addition, the cases with injected eccentricity $e_{0.01 \, \mathrm{Hz}}=0.1$ generally lead to weaker bounds than the corresponding circular injections, as expected from the degeneracy among $b$, $e_{0.01 \, \mathrm{Hz}}$, and $\mathcal{M}$ discussed above.
Despite this, most of the resulting constraints remain within the same order of magnitude, with typical upper bounds on $|b|$ at the level of $\lesssim\mathcal{O}(10^{-7})$.

One notable exception is the LISA-only case with $e_{0.01\,\mathrm{Hz}}=0.1$, where the bound weakens by about an order of magnitude.
Inspection of the posterior shows a multimodal structure in this case, which broadens the inferred distribution of $b$ substantially.
Weak traces of similar secondary support are also present in the TianQin-only samples, although they have little impact on the resulting constraint.
The corresponding full posterior is shown in Appendix~\ref{app}.
This behavior further illustrates that once eccentricity is present, the harmonic structure of the signal can turn the degeneracy among $b$, $e_{0.01 \, \mathrm{Hz}}$, and $\mathcal{M}$ into a multimodal posterior with multiple competing solutions, making the beyond-\ac{GR} interpretation more fragile.
The combined TianQin+LISA network suppresses this effect and recovers a substantially tighter constraint.
This suggests that the multimodal structure is likely associated with the underlying parameter degeneracies of the eccentric dipole waveform, while the detector response and \ac{SNR} determine how strongly secondary peaks appear.

It is also worth emphasizing that the relative performance of TianQin and LISA for an individual source is not always similar, but depends on parameters such as the sky location \cite{Wang2026}.
Nevertheless, the present results clearly demonstrate the constraint of space-based observations.
Although this is weaker than the optimistic estimate of $\lesssim \mathcal{O}(10^{-9})$ reported in previous studies \cite{Barausse:2016eii,Liu2020a,Toubiana2020a}, full Bayesian parameter estimation in Gaussian noise can already constrain $|b|$ to the $\lesssim \mathcal{O}(10^{-7})$ level for sources with individual \acp{SNR} of only $\sim 7.5$ and a conservative one-year observation.

\section{Summary and discussion\label{sec5}}

In this work, we have developed, for the first time, a Bayesian framework that simultaneously incorporates orbital eccentricity and dipole-radiation corrections in the multiband \ac{GW} analysis of \acp{sBBH}.
This is particularly important for theory-agnostic tests of dipole-like \ac{GW} radiation in the inspiral, where both eccentricity and beyond-\ac{GR} corrections enter at negative \ac{PN} order and accumulate in the same low-frequency part of the signal.
Our results show that eccentricity cannot be neglected if one aims to obtain robust constraints in such tests.
Ground-based observations tightly constrain most of the shared source parameters, while the space-based inspiral observation preserves the early-time information needed to probe eccentricity and dipole radiation.
Multiband observations are therefore powerful, but only if these coupled effects are modeled consistently rather than neglected.

Using a Gaussian-noise Bayesian pipeline with full posterior inference, we find that the dipole parameter $b$ is strongly degenerate with both the chirp mass and the eccentricity.
As a result, including eccentricity can broaden the inferred dipole parameter posterior by opening an additional degeneracy direction.
For the circular injection considered in Sec.\ref{sec4}, this effect increases the width of the credible interval of $b$ by a factor of approximately three relative to the restricted analysis in which eccentricity is fixed to zero.
Neglecting eccentricity can therefore lead to an apparent non-\ac{GR} signature, or at least to an overly optimistic estimate of how well dipole radiation can be constrained.
Despite the degeneracy, the multiband strategy retains significant constraining power.
For a GW231123-like source,  complemented with information from ground-based detectors, the typical space-based constraints remain at the level of $|b| \lesssim \mathcal{O}(10^{-7})$ for signals with individual \acp{SNR} of $\gtrsim 7.5$ in a conservative one-year observation.
The comparison among TianQin, LISA, and TianQin+LISA further shows that combining detectors generally improves the constraint and helps suppress eccentricity-induced multimodal posteriors.
More broadly, this multiband \ac{sBBH} channel should not be viewed simply as the same as existing dipole-radiation tests from pulsars \cite{Freire:2012mg, Shao:2017gwu, Zhao:2022vig} or GW170817-like binary neutron star mergers \cite{Abbott2019a}.
These probes involve different source classes (namely, black hole spacetime with singularity) and different dynamical regimes, and therefore constrain complementary aspects of possible beyond-\ac{GR} effects.
In this sense, multiband \ac{GW} observations of \acp{sBBH} provide a distinct test of theory-agnostic dipole-like radiation, especially because they probe the long, low-frequency inspiral directly.

The present analysis is intentionally simplified, but it is sufficient to reveal the main physical effect discussed here.
On the waveform side, we use the inspiral-only, nonspinning \texttt{EccentricFD} model, and the ground- and space-based analyses rely on different waveform families.
These choices are mainly driven by the current lack of a computationally practical model that can simultaneously describe eccentricity, spin and the full inspiral--merger--ringdown signal across both bands with long-duration evolution.
In addition, in some theories, black holes can carry scalar charges or nontrivial sensitivities that may depend on spin and affect the binary dynamics or scalar emission \cite{Almeida2024,Berti2021,Elley2022}.
The parameter $b$ used here should therefore be interpreted as a phenomenological dipole-radiation coefficient, rather than as a theory-specific function of the component masses and spins.
Quantifying the possible bias or overconstraint due to the omission of spin would require a validated waveform model in a given theory that consistently includes spin, eccentricity, and dipole-radiation corrections.
On the observational side, we assume a continuous one-year observation and do not include duty cycles or data gaps.
We also neglect the overlap of multiple signals in the space-based data.
These effects can affect the achievable constraints in realistic analyses, with an impact that depends on the gap structure, effective observing time, source population, and foreground-subtraction performance.
A full assessment of these effects is beyond the scope of this work.
Our results should therefore be interpreted as an idealized single-source analysis that isolates the intrinsic eccentricity-induced degeneracy.
The fact that this degeneracy is already significant in this simplified setup indicates that it is a robust feature of the underlying low-frequency phase structure that should be included in future, more realistic analyses.
This in turn suggests that future precision tests will require more complete waveform models and more realistic treatments of the data.
More generally, other negative-\ac{PN} effects may also enter the same low-frequency phase evolution, reinforcing the need for broad and self-consistent modeling in future multiband \ac{GW} tests of gravity.

The data supporting the findings of this study will be made publicly available at \url{https://github.com/HumphreyWang/eccentric_bbh_dipole_public}.

\section*{Acknowledgments}
We thank Sizheng Ma for helpful discussions, and the anonymous referee for comments.
This work is supported by the Beijing Natural Science Foundation (1242018), the National Natural Science Foundation of China (12573042), the National SKA Program of China (2020SKA0120300), the Max Planck Partner Group Program funded by the Max Planck Society, and the High-performance Computing Platform of Peking University.
H. Wang is supported by the Key Laboratory of TianQin Project (Sun Yat-sen University), Ministry of Education.

\bibliography{ref}

\appendix
\section{Full posterior example and multimodal structure\label{app}}

\begin{figure*}
	\centering
	\includegraphics[width=\linewidth]{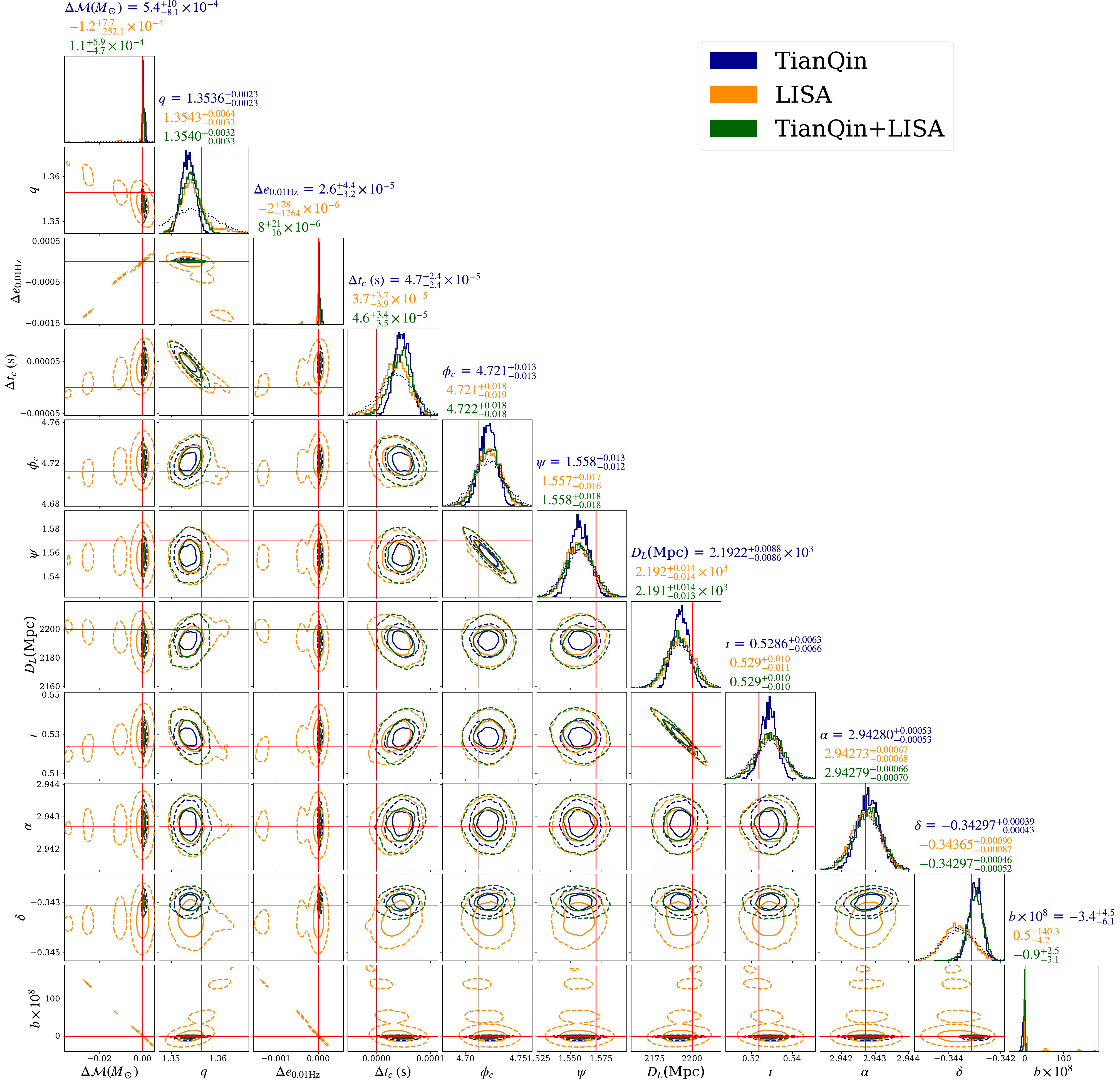}
	\caption{Full posterior distributions for the $e_{0.01 \, \mathrm{Hz}}=0.1$ injection with different detector configurations.
		Red lines indicate the injected values.
		The two-dimensional contours enclose 50\% and 90\% of the posterior probability, and the dashed curves in the one-dimensional panels show the corresponding prior distributions.}
	\label{pe_full}
\end{figure*}

In the main text, we focused on the four parameters most relevant to the dipole-radiation test, i.e. $\mathcal{M}$, $q$, $e_{0.01 \, \mathrm{Hz}}$, and $b$.
Here we present the full posterior distributions for the $e_{0.01 \, \mathrm{Hz}}=0.1$ injection in Fig.~\ref{pe_full}.
The most prominent nontrivial structure appears in the LISA-only analysis, where multiple posterior peaks emerge.
These peaks are arranged mainly along the degeneracy direction involving $\mathcal{M}$, $e_{0.01 \, \mathrm{Hz}}$, and $b$, providing a more explicit illustration of the correlation structure discussed in the main text.
The weaker bound on $b$ in the LISA-only configuration is therefore associated with multiple competing solutions in the full parameter space.
The TianQin-only posterior is dominated by the main peak, but weak traces of secondary support can still be seen in the samples.
These secondary peaks are suppressed in the combined TianQin+LISA analysis, indicating that the network helps break the degeneracies responsible for the multimodal structure.

Apart from the parameters highlighted in the main text, the remaining dimensions are generally dominated by the ground-informed priors.
Small residual shifts relative to the injected values mainly reflect the limited additional information provided by the space-based likelihood in those directions, together with fluctuations from the Gaussian noise realization.
Constraints on declination angle $\delta$ show modest differences among detectors because of their different orbital configurations.
These secondary variations do not affect the main conclusion of this work that the dominant nontrivial structure is the coupled degeneracy among $\mathcal{M}$, $e_{0.01 \, \mathrm{Hz}}$, and $b$.

\end{document}